\documentclass[aps,eqsecnum]{revtex4}
\usepackage{graphicx}
\def\beq{\begin{eqnarray}}
\def\eeq{\end{eqnarray}}
\def\R{{\cal R}}
\def\C{{\cal C}}
\def\M{{\cal M}}
\def\e{\alpha}
\def\ezm{{\e_{\rm zm}}} % zero mode
\def\ebs{{\e_{\rm bs}}} % bound state
\def\hezm{{\hat{\e}_{\rm zm}}} % zero mode
\def\hebs{{\hat{\e}_{\rm bs}}} % bound state
\renewcommand{\Re}{{\rm Re}}
\renewcommand{\Im}{{\rm Im}}
\begin{document}

\title{Massive scalar states localized   on a de Sitter brane}
\author{David Langlois}
\affiliation{GReCO,
Institut d'Astrophysique de Paris (CNRS),\\
98bis Boulevard Arago, 75014 Paris, France}
\author{Misao  Sasaki}
\affiliation{Department of Earth and Space Science,
Graduate School of Science,\\
Osaka University, Toyonaka~560-0043, Japan}

\begin{abstract}
We consider a brane scenario with a massive scalar field in the 
five-dimensional bulk. We study the scalar states that are localized on 
the brane, which is assumed to be de Sitter. These localized scalar modes 
are massive in general, their effective 
four-dimensional mass depending on the mass of the 
five-dimensional scalar field, on the Hubble parameter in the brane and on 
the coupling between the brane tension and the bulk scalar field. We then 
introduce a purely four-dimensional approach based on an effective 
potential for the projection of the scalar field in the  brane, and discuss 
its regime of validity. Finally, we explore the quasi-localized scalar 
states, which have a non-zero width that quantifies their probability of 
tunneling from the brane into the bulk. 
\end{abstract}

\maketitle

\section{Introduction}
A lot of attention has been recently devoted to the braneworld scenario,
which provides an alternative to the standard Kaluza-Klein compactification.
In fact, this idea was initiated a long time ago
 by the suggestion \cite{defect} that
particles could be localized on a defect embedded in a higher-dimensional
space, the simplest example being the case of a domain wall, or three-brane,
in a five-dimensional bulk spacetime. It was then shown, ignoring gravity,
 that one could localize scalars and fermions on the domain wall.

Recent progress was achieved on the gravitational aspect of the problem,
by showing that it was possible to localize massless gravitons \cite{rs99b}, 
and thus
to recover standard four-dimensional gravity, on a brane with tension
embedded in an AdS bulk, with the appropriate (negative) cosmological
constant
$\Lambda=-{6/\ell^2}$,
so that the brane is effectively Minkowski. In addition to the
massless gravitons, one finds a continuous spectrum of massive gravitons,
starting from zero mass but with a weak coupling to the brane so that their
contribution becomes important only on scales of the order of $\ell$ or
below.

The analysis of the graviton modes has also been extended to the case
of a  brane whose effective geometry is de Sitter \cite{gs99,lmw00}, 
which is the case when
the brane tension more than compensates the cosmological constant.
The mode analysis on this background yields  a massless graviton, analogous
to the four-dimensional graviton, plus a continuum of massive modes, which
starts at $(3/2)H$. There is thus a gap between the zero mode and the massive
modes. A recent work \cite{fk02} 
has generalized this treatment to the case of
conformally flat cosmological
geometries and shown that the minimum gap is $\sqrt{3/2} H$.

Not surprisingly, massless bulk scalar fields have properties very similar
to the gravitons. In the Randall-Sundrum (RS) background, it is 
easy to  show that,
as for gravitons, there exist a localized zero mode and a continuum of
arbitrary light states.
When one allows for a small but non-vanishing mass, it turns out that
the massless bound state disappears and turns into a quasi-localized
state with finite mass and finite width, the latter quantifying
the probability for the state to decay into the bulk \cite{drt00}.

The purpose of the present work is
 to study how these properties of massive bulk
scalar fields are modified when the brane geometry is de Sitter instead  
of Minkowski.
Massive scalar fields are  of special interest for
a bulk inflaton model of braneworld inflation
% the scenario of bulk inflation
\cite{hs00,shs01,hts01}, 
where inflation in the brane is driven by a bulk
scalar field, in contrast with
%brane inflation, 
a brane inflaton model,
where inflation is driven
by a four-dimensional field confined to the brane \cite{mwbh}.

In the present work, we study in detail the {\it bound states} of a
massive bulk scalar field, localized on a {\it de Sitter brane}, taking 
also into
account the possibility of a {\it coupling} between the scalar field and
the tension of the brane. We show that, depending on the range of values
for the scalar field mass $M$ and the coupling, one can find bound states,
which can be massless or massive.

We then consider and extend another approach, which consists in determining
an effective four-dimensional potential for the projection of the
bulk scalar field in the brane, taking into account the five dimensional
effects, via the effective four-dimensional Einstein equations. We are
thus able to deduce an effective four-dimensional potential that combines
the five-dimensional potential and the coupling to the brane.
We then compare this approach with the mode approach and find
surprisingly good agreement. Namely, given the five-dimensional mass,
both the critical coupling allowing for a  zero mode  and that
at which the bound-state ceases to exist agree well
with those obtained from the mode analysis.

Finally, we also explore some aspects of
the quasi-bound states (or quasi-normal modes) in the range 
of parameters where  the bound states
no longer  exist.

The plan of the paper is the following. In the next section, we present 
the model and give the equation of motion for the bulk scalar field. 
In section 3, we discuss the existence of bound states. Section 4 is 
devoted to an alternative approach based on an effective four-dimensional 
potential. Finally, we explore, analytically and 
 numerically, the quasi-localized 
scalar modes, which correspond to modes escaping from the brane into 
the bulk.

%%%%%%%%%%%%%%%%%%%%%%%%%%%%%%%%%%%
\section{Description of the model}
%%%%%%%%%%%%%%%%%%%%%%%%%%%%%%%%%%%
We consider a five-dimensional bulk scalar field $\phi$ and a brane
with a tension $\sigma$ coupled to this bulk scalar field. The corresponding
action is given by
\begin{eqnarray}
S=\int d^5x\sqrt{-g^{(5)}}
\left({1\over2\kappa^2}(R-2\Lambda_5)-{1\over2}(\nabla\phi)^2
-V(\phi)\right)
-\int d^4x\sqrt{-g^{(4)}}\sigma(\phi),
\end{eqnarray}
where we have introduced  a five-dimensional (negative) cosmological constant
\beq
\Lambda_5=-{6\over \ell^2}.
\eeq

In this work, we will concentrate, for simplicity,
 on the case of a quadratic
bulk potential,
\beq
V(\phi)={1\over 2}M^2\phi^2,
\label{V}
\eeq
and of a quadratic coupling, i.e. of the form
\beq
\sigma(\phi)=\sigma_0+{\e\over \ell}\phi^2,
\label{sigma}
\eeq
where $\e$ is a dimensionless parameter
characterizing the `strength' of the coupling
to the brane.

We will take as our  background configuration, the solution of the above
setup, with the usual cosmological symmetries (homogeneity and isotropy
along the three ordinary spatial dimensions), when
 the scalar field vanishes everywhere.
In this case, the bulk is effectively empty of matter and, owing to
the cosmological symmetries we impose, its geometry corresponds
to AdS$_5$. The metric can be written in the form
\beq
ds^2=dr^2+\ell^2 \sinh^2(r/\ell)\left[-d\tau^2+e^{2\tau}
d{\bf x}_{(3)}^{2}\right],
\label{ads_metric}
\eeq
where we have chosen, for simplicity,
a flat slicing for the three-dimensional
surfaces.

It is possible to insert a brane in such a geometry. Assuming as
usual that the bulk is $Z_2$ symmetric, i.e. mirror symmetric, about
the brane, the junction conditions for the metric at the brane location
yield the familiar brane Friedmann equation \cite{bdel99}
\beq
H^2={\Lambda_5\over6}+\left({\kappa^2\over6}\right)^2\sigma_0^2.
\eeq
In the present case, the Hubble parameter is constant, so that the
brane geometry is de Sitter. The Hubble parameter is related
to the brane position $r_0$, which is fixed in the above coordinate system
(\ref{ads_metric}), 
according to the expression
\begin{equation}
H^2={1\over\ell^2\sinh^2 (r_0 /\ell)}.
\label{H_r}
\end{equation}
It is then convenient to rescale the time coordinate so that it corresponds
to the cosmic time in the brane. The metric now reads
\begin{equation}
ds^{2}=dr^{2}+H^2\ell^2\sinh^2(r/\ell)
[-dt^{2}+H^{-2} e^{2Ht} d{\bf x}_{(3)}^{2}].
\label{metric}
\end{equation}
It is also convenient to introduce, in addition to the AdS lengthscale
$\ell$, another lengthscale, defined from the brane tension $\sigma_0$,
\beq
\ell_0=\left({\kappa^2\over6}\sigma_0\right)^{-1}.
\label{ellzero}
\eeq
In the RS case, $\ell=\ell_0$, whereas,
for a de Sitter brane, $\ell_0<\ell$.
The brane Hubble parameter is thus given by
\beq
H^2={1\over \ell_0^2}-{1\over \ell^2}.
\eeq
Introducing the variable
\beq
z=\cosh(r/\ell),
\label{z}
\eeq
the Hubble parameter given in (\ref{H_r}) can be reexpressed as
\beq
H(z_0)={1\over \ell\sqrt{z_0^2-1}}
\label{H_z}
\eeq
and the expression of $\ell_0$ in terms of $z_0$ is therefore
\beq
\ell_0={\ell\over z_0}\sqrt{z_0^2-1}.
\eeq

Let us now turn to  the scalar field, which will be considered as a
test field on the background configuration previously defined.
In other words, we will ignore its backreaction on the geometry.
In the bulk, the scalar field must satisfy the Klein-Gordon equation,
which is given, for the metric (\ref{metric})  by
\beq
{1\over{H^{2}\ell^{2}\sinh^{2}(r/\ell)}}
e^{-3Ht}\partial_{t}\left(e^{3Ht}\partial_{t}\phi\right)
-{1\over{\sinh^{4}(r/\ell)}}\partial_{r}\left(\sinh^{4}(r/\ell)
\partial_{r}\phi\right)+M^2\phi=0\,.
\label{bulkeq}
\eeq
In addition to this bulk equation, the scalar field must also satisfy the
appropriate boundary condition due to the presence of the brane. It can
be shown that, for $Z_2$ symmetry, this boundary condition is given by
\beq
\partial_r\phi|_{r=r_0}=-{1\over2}{d\sigma\over d\phi}
=-{\e\over \ell}\phi.
\label{boundary}
\eeq
It is clear from the two above equations (\ref{bulkeq}) and (\ref{boundary})
that we chose to work with quadratic expressions both for the bulk
potential and the brane coupling in order to get linear equations for the
scalar field, which greatly simplifies the analysis.

%%%%%%%%%%%%%%%%%%%%%%%%%%%%%%%%%%%%
\section{Existence of bound states}
%%%%%%%%%%%%%%%%%%%%%%%%%%%%%%%%%%%%
The purpose of this section is to solve explicitly the system of equations
(\ref{bulkeq}--\ref{boundary}). Since the bulk equation is separable,
it is natural to look for solutions of the form
\beq
\phi=u(r)\psi(t),
\eeq
which leads to a separation of the bulk differential
equation (\ref{bulkeq}) into a radial equation  and a time-dependent equation,
\begin{eqnarray}
&&\left[{1\over \sinh^{4}(r/\ell)}
\partial_{r}\left(\sinh^{4}(r/\ell)\partial_{r}\right)-M^2
+{\lambda^2\over \ell^2\sinh^{2}(r/\ell)}\right]u(r)=0\,,
\label{ueq}
\\
&&\left[{d^2\over dt^2}+3H{d\over dt}+H^2\lambda^2\right]\psi(t)=0\, ,
\label{psieq}
\end{eqnarray}
where $\lambda^2$ is a separation constant. And
the boundary condition (\ref{boundary}) becomes
\begin{eqnarray}
{d\over dr} u(r)=-{\e\over \ell} u(r)
\qquad\mbox{at}\quad~r=r_0\, .
\label{boundary2}
\end{eqnarray}

The solution of the radial equation (\ref{ueq}) can be expressed
 in terms of associated Legendre functions.
Using  the variable $z$ defined in (\ref{z}), one finds
\beq
u_\mu(z) = {P_{\nu-1/2}^{\mu}(z)
\over(z^2-1)^{3/4}} \,
\label{u}
\eeq
with
\beq
\mu^2={9\over 4}-\lambda^2, \qquad \nu=\sqrt{M^2\ell^2+4}.
\eeq
The effective four-dimensional mass of the mode, which appears 
clearly in the time-dependent equation (\ref{psieq}), is thus given by
\beq
m_{(4)}^2=\lambda^2H^2
=(9/4-\mu^2)H^2.
\eeq

In the rest of this section, we  wish to investigate whether there
exist bound state solutions.
Bound states modes are defined as solutions which are normalizable,
i.e., such that
\beq
2\int_0^{r_0} dr \sinh^2 r\  u_\mu(r) u_\mu^*(r)=
2\int_1^{z_0} dz \sqrt{z^2-1}\  u_\mu(z) u_\mu^*(z)<\infty.
\label{normalization}
\eeq
Since the behaviour of the Legendre functions when $z$ approaches $1$ is given
by 
\beq
P^\mu_\nu(z)\sim {2^{\mu/2}\over \Gamma(1-\mu)}(z-1)^{-\mu/2}, 
\qquad z\rightarrow 1,
\label{Pz1}
\eeq
the integral in (\ref{normalization}) converges if
$\mu$ is real and negative.
Therefore, a bound state exists if one can find  a solution (\ref{u}) with 
$\mu<0$
that  also satisfies the boundary condition (\ref{boundary2}), which can be
reexpressed in terms of the variable $z$ as
  \begin{eqnarray}
{d\over dz}u+{\e\over(z^2-1)^{1/2}}u=0\quad\mbox{at}~z=z_0\,.
\end{eqnarray}
Substituting the solution (\ref{u}), this gives the condition
\begin{eqnarray}
{d\over dz}P^\mu_{\nu-1/2}-{3\over2}{z\over z^2-1}P^\mu_{\nu-1/2}
+{\e\over(z^2-1)^{1/2}}P^\mu_{\nu-1/2}&=&\cr
{1\over z^2-1}
\left[\left((\nu-2)z +\e(z^2-1)^{1/2}\right)P^\mu_{\nu-1/2}
-(\mu+\nu-1/2)P^\mu_{\nu-3/2}\right]
&=&0\, ,
\label{junction}
\end{eqnarray}
where the second expression is obtained by using the recurrence relations
satisfied by the associated Legendre functions.

The bound states are delimited by two extreme cases: the massless mode
(zero mode),
characterized by the value $\mu=-3/2$, and the mode at the top of the gap,
characterized by the mass-squared $m_{(4)}^2=(3/2)H^2$, i.e.,
$\mu=0$, above which bound states cannot exist.
For each set of values for  $H$ (or $z$) and $M$, one can thus define
two critical values for the coupling, the lower one, $\ezm$, corresponding
to a bound state that is also a zero mode and the upper one, $\ebs$
giving the threshold above which  bound states disappear.

Using the boundary condition (\ref{junction}), one immediately finds
that the zero mode coupling, corresponding to $\mu=-3/2$, is given by
\beq
\ezm(z,M)
=-M^2\ell^2\; {P^{-5/2}_{\nu-1/2}\over P^{-3/2}_{\nu-1/2}}\,.
\eeq
where we have used one of the  recurrence relations for the Legendre functions 
and the definition of $\nu$ in terms of $M$.
 Note that, for $M=0$, one gets  $\ezm=0$:
one recovers the usual zero mode of a massless bulk scalar field
with  no coupling to the brane. As soon as the bulk mass $M$ is nonzero,
with $M^2>0$,
the zero mode coupling becomes negative,
which means that one can find a zero
mode localized on the brane only with a negative coupling.
The case opposite to this
occurs for an unstable potential, i.e., $M^2<0$ (for
$M^2\ell^2>-4$, which we assume in the present paper).

It is interesting to consider two limits concerning the Hubble parameter
of the dS brane: the limit where the Hubble parameter
is very small, $H\ell\ll 1$, corresponding to large values of $z$; and, 
conversely, the limit where $H\ell \gg 1$, corresponding to values of $z$
very close to $1$.
For the first limit, one can use the asymptotic behaviour 
\beq
P^\mu_\nu(z)\sim 2^\nu\pi^{-1/2}{\Gamma\left(\nu+{1\over 2}\right)
\over \Gamma\left(1+\nu-\mu\right)} z^\nu, \qquad z\gg 1,
\label{Pzinf}
\eeq
which leads to the result 
\beq
\ezm \simeq 2-\nu= 2-\sqrt{M^2\ell^2+4}, \qquad
 H\ell \ll 1.
\eeq
For the second limit, the expression (\ref{Pz1}) shows 
 that $\ezm$ behaves like the inverse of the Hubble parameter, more
specifically 
\beq
\ezm \simeq -{M^2\ell\over 5H}, \qquad H\ell \gg 1.
\label{ezmz1}
\eeq

\begin{figure}
\begin{center}
\includegraphics[width=4.8in]{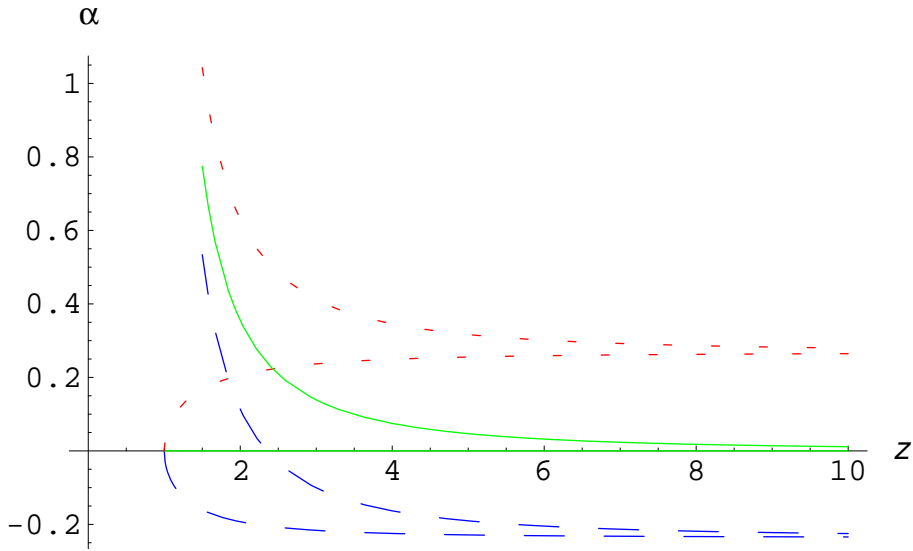}
\end{center}
\caption{Range of coupling values $\e$ that allow for the existence of 
a bound state, as a function of $z$ (and thus of $H\ell$) 
 for the three following cases: 
$M^2\ell^2=1$ (long-dashed blue lines);  $M^2\ell^2=-1$ 
(short-dashed red lines); $M^2\ell^2=0$ (continuous green lines).
In each case the upper limit corresponds to $\ebs(z,M)$ and the 
lower limit to $\ezm(z,M)$, both  values converging towards 
$\e=2-\sqrt{4+M^2\ell^2}$ at large $z$.  }
\label{epsilon}
\end{figure}

Similarly, one can define the largest coupling, $\ebs$, that allows 
for   the existence of
a bound state. 
Using  the boundary condition (\ref{junction}) with $\mu=0$, it is
 given by the expression
\beq
\ebs(z,M)= {(\nu-1/2)P_{\nu-3/2}- (\nu-2)z P_{\nu-1/2}
\over (z^2-1)^{1/2}P_{\nu-1/2}}.
\eeq
When the mass of the bulk scalar field vanishes, $M=0$, then $\nu=2$ and
the above expression for the critical coupling  simplifies and reads
\beq
\ebs(z,0)= {3\over 2}{P_{1/2}\over (z^2-1)^{1/2}P_{3/2}}.
\eeq
As before, it is interesting to consider the behaviour of 
the maximal coupling in the two limits $z\rightarrow\infty$ and 
$z\rightarrow 1$.
In the first limit, one finds the same result as for $\ezm$, namely
\beq
\ebs\simeq  2-\nu= 2-\sqrt{M^2\ell^2+4},  \qquad
 H\ell \ll 1,
\eeq
which means that a bound state exists only for a value of the coupling
very close to $2-\nu$, i.e., negative for $M^2>0$ (and positive otherwise).
In the second limit, one finds that $\ebs$ is proportional to $H$,
\beq
\ebs\simeq {3\over 2} H\ell, \qquad H\ell \gg 1.
\eeq
In Fig. \ref{epsilon}, we have plotted the critical couplings 
$\ezm$ and $\ebs$ as a function of $z$ for three bulk masses ($M^2\ell^2=1$,
$M^2\ell^2=0$, $M^2\ell^2=-1$). This defines, for the three cases, the 
regions in  the coupling parameter space that allow for a bound state. Outside
these regions, there is no bound state and the four-dimensional 
mass of the modes has an imaginary part, as will be discussed later.
For large $z$, i.e. for small $H$, the bound state regions become thinner
and thinner bands around the line $\e=2-\nu$.
  
Note that, for a positive  mass $M$, 
the critical coupling $\ebs$ is not always positive, but
becomes negative for $z>z_c(M)$. This means that whereas
for $z< z_c(M)$, i.e., for
a Hubble parameter sufficiently high, one can always find a bound state with
a vanishing coupling, it is no longer the 
case for $z>z_c(M)$ where  a bound state can be found only 
with a negative coupling.

%%%%%%%%%%%%%%%%%%%%%%%%%%%%%%%%%%%%%%%%%%%%%%%%%%%%%%%

%%%%%%%%%%%%%%%%%%%%%%%%%%%%%%%%%%%%%%%%%%%%%%%%%%%%%%%%%

Finally, a more general quantitative analysis is possible in the 
cases where  $M^2\ell^2\ll 1$ and $\e\ll 1$. Indeed, 
by  linearizing  the boundary condition (\ref{junction}) 
about the zero mode solution characterized by  $\mu=-3/2$, $M=0$ and 
$\e=0$, one can find the linear deviation $\Delta\mu=\mu+3/2$ as 
a function of the small parameters $M^2\ell^2$ and $\e$.
Expressing $\Delta\mu$   in terms of the 
effective four-dimensional mass of the bound state, one gets
\beq
m_{(4)}^2 \simeq f_1(H\ell) {M^2\over 2}+ f_2(H\ell) {2\e\over \ell^2},
\label{linear}
\eeq
with
\beq
f_1(H\ell)={\sqrt{1+H^{2}\ell^{2}}\over\R(H\ell)}
-{3\over2}H^2\ell^2\,,
\quad f_2(H\ell)={1\over\R(H\ell)}, 
\eeq
and
\beq
\R(H\ell)={2\sqrt{z^2-1}\,P^{-3/2}_{1/2}(z)\over 3P^{-3/2}_{3/2}(z)}
=\sqrt{1 + H^2\,\ell^2} 
- H^2\,\ell^2\,\ln \left(\frac{1 + {\sqrt{1 + H^2\,\ell^2}}}{H\,\ell}\right).
\eeq
The expression (\ref{linear}) shows that, in the linearized limit,
the (small) bulk mass-squared and 
brane coupling contribute additively to the effective four-dimensional 
mass-squared of the
bound state. The coefficients in the  linear combination depend on $z$, i.e.
on the value of the Hubble parameter. 
In the limit $H\ell\rightarrow 0$, i.e., $z\rightarrow +\infty$,
one finds
\beq
m_{(4)}^2 \simeq {M^2\over 2}
+2 {\e\over \ell^2}, \qquad H\ell\ll 1,
\label{linear1}
\eeq
whereas in the opposite limit one gets
\beq
m_{(4)}^2 \simeq {3\over 5}{M^2}
+3 {H\over \ell} \e, \qquad H\ell \gg 1. 
\label{linear2}
\eeq
The contribution from the coupling is thus proportional to the coupling 
parameter $m$ times the largest of the two mass scales $H$ and $\ell^{-1}$.
The contribution from the bulk scalar field mass is essentially the same
with a very small variation of the coefficient. Note also that the result 
(\ref{ezmz1}) follows immediately from (\ref{linear2}) with 
$m_{(4)}^2=0$.

In the case of $H\ell\ll1$, it may be worth mentioning that
the coupling $(\e/\ell)\phi^2$ on the brane contributes to the effective mass
term as it is, if we rescale $\phi$ as $\phi\to\Phi=\sqrt{\ell}\,\phi$
by introducing an effective four-dimensional scalar field $\Phi$
of correct dimensions, whereas the bulk mass-squared $M^2$ 
contributes with a factor of $1/2$.

%%%%%%%%%%%%%%%%%%%%%%%%%%%%%%%%%%%%%%%%%%%
\section{Effective potential approach}
%%%%%%%%%%%%%%%%%%%%%%%%%%%%%%%%%%%%%%%%%%%

In this section, we will use the four-dimensional projection
of the five-dimensional Einstein equations onto the brane
in order to establish an effective 
potential for the four-dimensional projection of the scalar field on the
brane.
Such a procedure was successfully applied in the case of a bulk scalar field
with quadratic potential but without coupling to the brane \cite{shs01,hts01} 
and it was shown that the effective potential is simply half
the bulk potential when $H\ell\ll1$. The purpose
of this section is to generalize this result to include the coupling of
the scalar field to the brane.

%%%%%%%%%%%%%%%%%%%%%%%%%%%%%%%%%
\subsection{General derivation}
%%%%%%%%%%%%%%%%%%%%%%%%%%%%%%%%%
We will start by recalling some results derived in \cite{mw00}
 following the procedure of \cite{sms99}. It was shown there 
in particular that  the effective four-dimensional
Einstein equations for  dilaton-vacuum configurations are given by
 \begin{eqnarray}
{}^{(4)}G_{\mu\nu}=
{2\kappa_5^{~2}\over 3} \hat{T}_{\mu\nu}(\phi)
-{}^{(4)}\Lambda g_{\mu\nu} - E_{\mu\nu}\,,
\label{einstein}
\end{eqnarray}
with
\begin{eqnarray}
\hat{T}_{\mu\nu}&=& D_\mu \phi D_\nu \phi -{ 5 \over 8} g_{\mu\nu}
(D\phi)^2
\,,
\label{4d-dilaton}\\
{}^{(4)}\Lambda &=&{{}^{(5)}\Lambda\over 2}+\frac{1}{2}\kappa_5^2
\left[V(\phi) +\frac{1}{6}\kappa_5^2\,\sigma^2
-{1 \over 8} \left({d\sigma \over d \phi}
\right)^2\right] \,,
\label{Lambda4}
\end{eqnarray}
where $D_\mu$ denotes the covariant differentiation with respect to
the metric on the brane.
Using the four-dimensional Bianchi identities, 
the covariant differentiation of 
Einstein's equations (\ref{einstein}) implies
\beq
D^\mu E_{\mu\nu}={2\over 3}\kappa_5^2 D^\mu \hat{T}_{\mu\nu}
-D_\nu {}^{(4)}\Lambda.
\label{DE}
\eeq

Specializing  to a FLRW (Friedmann-Lema\i tre-Robertson-Walker) 
 geometry, one finds
that the Friedmann equation on the brane, corresponding to the component
(0-0) of (\ref{einstein}) is given by
\beq
3H^2
=-{3\over\ell^2}
+\kappa_5^2\left({1\over4}\dot\phi^2+{1\over2}V+{\kappa_5^2\over12}\sigma^2
-{1\over16}\sigma'{}^2\right)+E,
\label{friedmann}
\eeq
where $E=E^0_0=-E_{00}$.  Equation (\ref{DE}) yields
\begin{eqnarray}
\dot E+4HE &=&
{\kappa_5^2}
\left[-{1\over2}\ddot\phi-2H\dot\phi
-{1\over2}V'-{\kappa_5^2\over12}(\sigma^2)'
+{1\over16}(\sigma'{}^2)'\right]\dot\phi\nonumber\\
&=&
{\kappa_5^2}
\left[{1\over2}\left(\ddot\phi+2H\dot\phi\right)
-\left(\ddot\phi+3H\dot\phi\right)
-{1\over2}V'-{\kappa_5^2\over12}(\sigma^2)'
+{1\over16}(\sigma'{}^2)'\right]\dot\phi,
\label{Eequation}
\end{eqnarray}
where we have rewritten the terms in the brackets involving the first and
second derivatives of $\phi$, as two linear combinations, one which will be
easily integrated and the other one corresponding to the familiar
four-dimensional Klein-Gordon equation.

We are now going to assume that
the brane value of the scalar field satisfies a Klein-Gordon equation
that  can be written in the form
\begin{eqnarray}
\ddot\phi+3H\dot\phi+V_{\rm eff}'=-J\,,
\label{effeq}
\end{eqnarray}
where $V_{\rm eff}$ is an effective potential which we wish to determine and
where $J$ stands for a possible energy leak out of the brane into the bulk.
Substituting this Klein-Gordon equation in Eq.~(\ref{Eequation}) above, one
finds
\begin{eqnarray}
\dot E+4HE
={\kappa_5^2\over2}\left(\ddot\phi+2H\dot\phi\right)\dot\phi
+\kappa_5^2\left[
V_{\rm eff}'-{1\over2}V'-{\kappa_5^2\over12}(\sigma^2)'
+{1\over16}(\sigma'{}^2)'+J\right]\dot\phi\,.
\label{Eeq}
\end{eqnarray}
This strongly suggests that the effective potential, if it makes sense, is
of the form
\beq
V_{\rm eff}=-{3\over\kappa_5^2\ell^2} +{1\over2}V
+{\kappa_5^2\over12}\sigma^2-{1\over16}\sigma'{}^2,
\label{Veff}
\eeq
where the choice of the constant is for convenience
as will be seen very soon.
Remarkably, this is the same combination which appears in the Friedmann
equation (\ref{friedmann}). That is, if we adopt this definition for 
the effective potential, and  introduce the quantity $X$ defined by 
\beq
E={\kappa_5^2\over4}\dot\phi^2+\kappa_5^2X,
\eeq
the Friedmann equation (\ref{friedmann}) and the equation (\ref{Eeq}) 
for $E$ reduce 
to the very simple system
\begin{eqnarray}
3H^2&=&\kappa_5^2
\left[{1\over2}\dot\phi^2+V_{\rm eff}+X\right]\,,
\label{efffried}
\\
\dot X+4HX&=&J\dot\phi,
\end{eqnarray}
where one recognizes the standard four-dimensional 
Friedmann equation with a scalar field and some extra component $X$.
The second equation is a  (non)-conservation equation for the extra-component
$X$. Note that $X$ plays the role of the Weyl, or
dark radiation, which was identified in the simplest model of brane cosmology
\cite{bdel99}. When the energy outflow is zero, i.e. $J=0$, one recovers
the result 
\beq
X={{\cal C}\over a^4}.
\eeq
A non zero $J$ means that there is an energy outflow from the brane into
the bulk, which is going to feed the Weyl energy density. This is similar 
somehow to the growth of ${\cal C}$ 
induced  by the gravitational wave emission 
from brane cosmological perturbations,  
which  was  recently analysed in \cite{lsr02}. 
 It should also be mentioned that we have not really proved the form 
(\ref{Veff}) for the effective potential. We have simply shown that 
this form  is consistent with the system of effective equations in the brane.

%%%%%%%%%%%%%%%%%%%%%%%%%%%%%%%
\subsection{Exact solutions}
%%%%%%%%%%%%%%%%%%%%%%%%%%%%%%%%
It is instructive at this stage to check the above approach for known 
exact solutions. One of the simplest examples is the case of a bulk scalar 
field with the exponential potential~\cite{static,cr99,LanRod01,Langlois02,KoyTak03}
\beq
V(\phi)=V_0\exp\left(-{2\over \sqrt{3}}
\lambda \kappa\phi\right).
\eeq
One can find for the bulk with vanishing cosmological constant 
explicit static solutions, which read
\beq
ds^2=-h(R)dT^2+{R^{2\lambda^2}\over h(R)}dR^2+R^2 d{\vec x}^2,
\label{dil_stat}
\eeq
for the metric, 
with 
\beq
h(R)=-{\kappa^2 V_0 /6\over 1-(\lambda^2/4)}R^2-\C R^{\lambda^2-2},
\eeq
 $\C$ being an arbitrary constant, and 
\beq
{\kappa\over \sqrt{3}}\phi=\lambda\ln(R).
\eeq
 for the scalar field. 

A brane with a tension 
\beq
\sigma(\phi)= \sigma_0 \exp\left(-{\lambda\over \sqrt{3}}
 \kappa\phi\right).
\eeq
will undergo a cosmological evolution governed by the generalized 
Friedmann equation 
\beq
H^2=\left[
{\kappa^4\over 36}\sigma_0^2
+{\kappa^2 V_0 /6\over 1-(\lambda^2/4)}\right] R^{-2\lambda^2}
+\C R^{-4-\lambda^2}.
\label{fried_cr}
\eeq
Since the bulk cosmological constant is zero, the effective potential is here
\beq
V_{\rm eff}={1\over2}V
+{\kappa_5^2\over12}\sigma^2-{1\over16}\sigma'{}^2
&=&
\left[{V_0\over 2}+{\kappa^2\over 12}\left( 1-{\lambda^2\over 4}\right)
\sigma_0^2\right]\exp\left(-{2\over \sqrt{3}}\lambda \kappa\phi\right)
\nonumber\\
&\equiv&
V_{\rm eff,0}\,\exp\left(-{2\over \sqrt{3}}\lambda \kappa\phi\right).
\eeq
It is not difficult to check that $\phi$ satisfies the effective 
Klein-Gordon equation (\ref{effeq}) with
\beq
J=-\left(1-{\lambda^2\over 2}\right)H\dot\phi\,.
\eeq
The Friedmann equation (\ref{fried_cr}) can also be written in the effective form 
(\ref{efffried}) 
with the  energy density $X$ given by
\beq
{\kappa^2}X
&=&
3\left( 1-{\lambda^2\over 4}\right)\C R^{-4-\lambda^2}
-{\kappa^2\over 4}\dot\phi^2
\nonumber\\
&=&3\left(1-{\lambda^2\over2}\right)\C R^{-4-\lambda^2}
-{\lambda^2/4\over1-\lambda^2/4}V_{\rm eff,0}\,R^{-2\lambda^2}.
\eeq
Thus, if we apply our interpretation of $J$ that it describes energy flow
from the brane  into  the bulk, the above analysis implies that the energy 
is actually flowing onto the brane for $\lambda^2<2$ rather than 
flowing out of the brane. It is also worth noting that, whereas the present 
exemple is an instructive check, 
it is however not very useful in practice because the energy 
exchange between the brane and the bulk is important so that the choice
between the variables $X$ and $E$ is somewhat arbitrary (in fact, the 
expression for $E$ in terms of the scale factor is simpler). It is 
really for the cases where the energy leak $J$ is small that the 
effective potential approach makes really sense physically.

\subsection{Quadratic case}
In the case of the quadratic bulk potential and brane coupling, 
(\ref{V}-\ref{sigma}),
the effective potential suggested by the above analysis reads
(with Eq.~(\ref{ellzero}) in mind),
\beq
V_{\rm eff}={3\over\kappa_5^2}
\left({1\over\ell_0^2}-{1\over\ell^2}\right)
+{1\over2}\left({1\over2}M^2
+2{\e\over \ell\ell_0}-{\e^2\over 2\ell^2}\right)\phi^2
+{\kappa_5^2\over 12}{\e^2\over\ell^2}\phi^4.
\eeq
At the extremum $\phi=0$, the effective mass-squared is thus given by
\beq
\M_{\rm eff}^2={1\over 2}M^2+2{\e\over \ell^2}\sqrt{1+H^2\ell^2}-
{\e^2\over 2\ell^2},
\label{Meff}
\eeq
where we have replaced $\ell_0$ by its expression in terms of $\ell$ and $H$.
It is interesting to note that this potential, which is quartic in $\phi$,
takes a double-well form for $\M_{\rm eff}^2<0$.
If the present approach is valid, this implies that
we may describe a situation of spontaneous symmetry breaking
with $V_{\rm eff}$.

We now wish to compare the results of the mode analysis with the effective
potential approach, and in order to so, to compare the effective mass
predicted by the two analyses.
In the previous section, we have identified two critical cases: the case
when the bound state is a zero mode, i.e., $m_{(4)}^2=0$, and the case
when the bound state reaches the top of the gap,
i.e., $m_{(4)}^2=(9/4)H^2$.
In the effective potential approach, the equation
\beq
\M_{\rm eff}^2\equiv{2\e\over \ell\ell_0}-{\e^2\over 2\ell^2}+{M^2\over 2}=0,
\eeq
is easily solve to yield
\beq
{\hezm}(z,M)={2z\over \sqrt{z^2-1}}\pm \sqrt{{4z^2\over {z^2-1}}+M^2\ell^2}.
\eeq
Similarly, the bound state threshold is determined by solving the equation
\beq
\M_{\rm eff}^2\equiv{2\e\over \ell\ell_0}-{\e^2\over 2\ell^2}+{M^2\over 2}=
{9/4\over \ell^2(z^2-1)},
\eeq
which gives
\beq
{\hebs}(z,M)={2z\over \sqrt{z^2-1}}\pm \sqrt{{4z^2\over {z^2-1}}+M^2\ell^2
-{9/2\over (z^2-1)}}.
\eeq
In both cases, we will keep only
the root with the minus sign since it matches
with the small coupling limit.
In the small Hubble parameter limit, i.e., in the large $z$ limit,
the two expressions converge towards the same value $2-\nu$ as found
previously.

For small bulk mass and coupling, the effective potential approach can 
provide a reasonable approximation, even in the case $H\ell\gg 1$.
Indeed, neglecting the $\e^2$ term  in (\ref{Meff}), one gets
\beq
\M_{\rm eff}^2\simeq {1\over 2}M^2+2{\e\over \ell^2}\sqrt{1+H^2\ell^2}, \qquad
(\e\ll M\ell).
\eeq
Therefore, in the limit $H\ell\ll 1$, one finds
\beq
\M_{\rm eff}^2\simeq {1\over 2}M^2+2{\e\over \ell^2}, \qquad H\ell\ll 1,
\eeq
which is exactly the same result as in (\ref{linear1}). 
This strongly indicates the validity of the effective potential approach
for $H\ell\ll1$,
at least in the case of quadratic potential and brane-coupling.
It is then tempting to conjecture that this approach is valid
for more general cases, including the case when the backreaction
of the scalar field dynamics to the geometry is non-negligible.

Even in the opposite limit $H\ell\gg 1$, one finds
\beq
\M_{\rm eff}^2\simeq {1\over 2}M^2+2{H\over \ell} \e, \qquad H\ell\gg 1,
\eeq
which is qualitatively similar to (\ref{linear2}) 
although the numerical coefficients
in the linear combination are now slightly different.

%%%%%%%%%%%%%%%%%%%%%%%%%%%
\section{Quasi-normal modes}
%%%%%%%%%%%%%%%%%%%%%%%%%%%
We have so far concentrated our attention on stable, bound-state modes.
It is however instructive to study as well the decaying, quasi-normal
modes in the case where there exists no bound-state mode,
although this is a more complicated problem.
In particular, if the effective potential we derived
in the previous section is valid, the term $J$ which describes the
possible energy leak to the bulk may be determined by studying the
decay width of quasi-normal modes.

Mathematically, quasi-normal modes are defined as those that satisfy
the purely outgoing-wave boundary condition at future Cauchy horizon
of AdS$_5$. 
A quasi-normal mode can be found by solving Eq.~(\ref{junction}) 
for complex $\mu$ with ${\rm Re}[\mu]>0$, with ${\rm Im}[\mu^2]$
being related  to the decay width of the effective 
four-dimensional mass $m_{(4)}$.
Note that, if $\mu$ is a solution to Eq.~(\ref{junction}), so is
its complex conjugate $\mu^*$.
For definiteness, let us reserve the terminology `quasi-bound-state modes'
for those modes with non-zero imaginary part, and call the modes
with positive real $\mu$ the `purely decaying modes'.

\subsection{Analytical approach}

In the limit  $H^2\ell^2\ll 1$,
one can obtain an analytic expression for
the quasi-bound-state modes with least real part, i.e., the 
effective mass with smallest decay width.
In this case, it is convenient to use one of the decompositions 
of the associated Legendre function into hypergeometric functions
\cite{Bateman}: 
\beq
P^\mu_\nu(z)=A_1 F\left({1\over 2}+{\nu\over 2}-{\mu\over 2}, 
{1\over 2}+{\nu\over 2}+{\mu\over 2}; \nu+{3\over 2},{1\over 1-z^2}\right)
+ A_2
F\left(-{\nu\over 2}+{\mu\over 2}, 
-{\nu\over 2}-{\mu\over 2}; -\nu+{1\over 2},{1\over 1-z^2}\right),
\eeq
with 
\beq
A_1={2^{-\nu-1}\,\Gamma\left(-{1\over 2}-\nu\right)
\over\sqrt{\pi}\,\Gamma\left(-\nu-\mu \right)}(z^2-1)^{-(\nu+1)/2}\,,
\quad
A_2={2^{\nu}\,\Gamma\left({1\over 2}+\nu\right)
\over\sqrt{\pi}\,\Gamma\left(1+\nu-\mu \right)}(z^2-1)^{\nu/2}\,.
\eeq 
Substituting this expression into the junction condition (\ref{junction}),
%\beq
%\left((\nu-2)z +\e(z^2-1)^{1/2}\right)P^\mu_{\nu-1/2}
%-(\mu+\nu-1/2)P^\mu_{\nu-3/2}=0,
%\eeq
and using
$z^2-1=(H\ell)^{-2}$,
one finds, by expanding in terms of $H^2\ell^2$,
\begin{eqnarray}
(\nu-2+\e)(\nu-1)
&+&\left\{{1\over 2}(\nu-2)(\nu-1)+
{1\over 4}\left[\left({1\over 2}-\nu\right)^2-\mu^2\right](\nu-4+\e)\right\}
H^2\ell^2
+{\cal O}(H^4\ell^4)
\cr
&+&2^{-2\nu}{ \Gamma(-\nu)\Gamma\left({1\over 2}+\nu-\mu \right)
\over\Gamma(\nu)\Gamma\left({1\over 2}-\nu-\mu \right)}
(1-\nu)(\nu+2-\e)(H^2\ell^2)^{\nu}
+{\cal O}\left((H^2\ell^2)^{\nu+2}\right)=0\,.
\label{expansion}
\end{eqnarray}
We stress that this expansion makes sense only for $\mu^2 H^2\ell^2\ll 1$.
The dominant terms are the first two terms which give
\beq
\mu^2 H^2\ell^2\simeq \left({1\over 2}-\nu\right)^2H^2\ell^2+ 
4{(\nu-2+\e)(\nu-1)\over \nu-4+\e}.
\eeq
For this  equation to be consistent
with the condition $\mu^2 H^2\ell^2\ll 1$, one must assume $\nu-2+\e\ll 1$,
which means that one must be very close to the bound state region defined 
before.
 In this case, 
the solutions  are 
\beq
\mu\simeq \pm\sqrt{\left(\e-{3\over 2}\right)^2-2{(1-\e)(\nu-2+\e)\over 
H^2\ell^2}}.
\label{lowestmu}
\eeq
In the more particular case where both  $M^2\ell^2\ll 1$ and $\e\ll1$, one
finds
\beq
\mu\simeq \pm\sqrt{{9\over 4}-{M^2\ell^2+4\,\e\over 2\,H^2\ell^2}}\,.
\eeq
It may be worth mentioning that this result is valid irrespective of
the relative magnitudes of $H\ell$, $M\ell$ and $\e$, as long as
they are all small compared to unity. In particular, if
the quantity inside the square root is positive, the solution with 
the minus sign (i.e., $\mu<0$) is just a bound-state solution
in the limit $H\ell\ll1$, given by Eq.~(\ref{linear1}).
On the other hand, if the quantity inside the square root
is negative, the dominant term of $\mu$ will become purely  imaginary
and a small real part will develop in $\mu$, as a result of
the appearance of an imaginary part in $\mu^2$.
The imaginary part of $\mu^2$, however, cannot be determined from the 
first line of (\ref{expansion}) which contains only polynomial functions 
of $\mu^2$. One must resort to the second line to compute its imaginary
part, which will be given by 
\beq
H^2\ell^2\Im{\left[\mu^2\right]}\simeq
{2^{2-2\nu}(1-\nu)(\nu+2-\e)\left(H\ell\right)^{2\nu}
\over (\nu-4+\e)}\Im{\left[
{ \Gamma(-\nu)\Gamma\left({1\over 2}+\nu-\mu \right)
\over\Gamma(\nu)\Gamma\left({1\over 2}-\nu-\mu \right)}\right]},
\label{imaginary}
\eeq
where one can substitute the solution (\ref{lowestmu}) in the right 
hand side. This expression  agrees with \cite{hts01} for $\e=0$. 

In the limit $H^2\ell^2\ll M^2\ell^2\ll1$ and $H^2\ell^2\ll\e\ll1$, 
one finds 
\beq
H^2\ell^2\Im{\left[\mu^2\right]}\simeq
 \pm{\pi\over 16}\left(M^2\ell^2+4\e\right)^2,
\label{mu2im}
\eeq
which agrees with the Minkowski limit of \cite{drt00} in the case $\e=0$.

Another case which may be studied analytically, though only
qualitatively, is when $\e\gg1$.
In this case Eq.~(\ref{junction}) implies that the terms
proportional to $\e$ should become $O(1/\e)$.
Hence, in the limit $\e\to\infty$, 
$\mu$ is given by a zero of $P^\mu_{\nu-1/2}(z)$ on the
complex $\mu$-plane (with $\Re[\mu]>0$ and $\Im[\mu]\neq0$), 
which is independent of $\e$. 
Thus $\mu$ converges to a finite value in the limit
$\e\to\infty$.

Let us next discuss the purely decaying modes for which
$\mu$ lies on the (positive) real axis, i.e., 
with no real part in the effective mass.
They are usually subdominant in the sense that their decay widths
are larger than the decay width of the complex quasi-bound-state
modes discussed above. 
For $H\ell\ll 1$, the asymptotic behaviour of Legendre functions is given 
in (\ref{Pzinf}) and one thus finds the 
 zeros of  the boundary condition (\ref{junction}) when 
the common dominator $\Gamma(1-\mu)$ for all terms goes to infinity, i.e. 
for 
\beq
\mu_n =1+n, \quad n=0,1,2, \dots.
\eeq
The other limit is $H\ell \gg 1$ and  the Legendre functions behave according
to (\ref{Pz1}).
In the boundary condition (\ref{junction}), the dominant terms, when $z\gg1$
are the terms proportional to $P^\mu_{\nu-1/2}$ and therefore the zeros 
of (\ref{junction}) are given by 
\beq
\mu_n={1\over 2}+\sqrt{4+M^2}+n, \quad n=0,1,2 \dots
\eeq
where $n$ is a positive integer, corresponding to cases where the Gamma
function in the denominator of $P^\mu_{\nu-1/2}$ blows up.
These zeros thus now depend  on $M$ but not on the coupling $\e$ or 
on $z$ (provided however $z\gg1$).

\subsection{Numerical approach}

In order to study the quasi-normal modes in more details,
we have solved numerically the following equation for the complex
number $\mu$,
\beq
F(\mu; M,\e,z)\equiv
-(\mu+\nu-1/2){P^\mu_{\nu-3/2}\over P^\mu_{\nu-1/2}}+(\nu-2)z+
\e(z^2-1)^{1/2}=0\,,
\eeq
for various values of the parameters $M$, $\e$ and $z$. 
As mentioned earlier, if $\mu$ is a solution, so is $\mu^*$.
Hence, without loss of generality, we may confine our search for
$\mu$ on the first quadrant (i.e., ${\rm Re}[\mu]>0$ and
${\rm Im}[\mu]>0$) on the complex $\mu$-plane.

Let us start our numerical exploration with the case of a vanishing
coupling $\e=0$.
%%%%%%%%%%%%%%%%%%%%%%%%%%%%%%%%%%%%%%%%%%%%%%%%%%%%%%%%%%%%%%%%%%%%%%%%%
 \begin{figure}
\begin{center}
\includegraphics[width=4.8in]{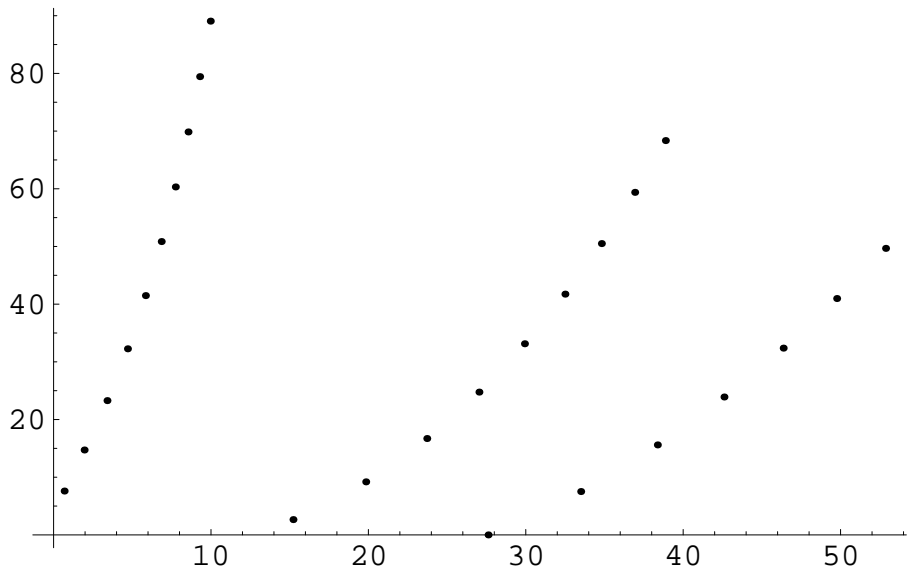}
\end{center}
\caption{Evolution  in the complex $\mu$-plane of the quasi-normal
modes when the mass $M$ varies (with $\e=0$, $z=10$).
 The highest points for each branch correspond to $M\ell=10$. The increment
between two adjacent points is $\Delta (M\ell)=1$.}
\label{qnm1}
\end{figure}
%%%%%%%%%%%%%%%%%%%%%%%%%%%%%%%%%%%%%%%%%%%%%%%%%%%%%%%%%%%%%%%%%%%%%%%%%
In Fig.~\ref{qnm1}, we have plotted  the numerical solutions 
 for $\mu$ in the complex plane for various values of the bulk mass $M$.
As can be seen in the figure, we have obtained several branches of solutions
which evolve continously as $M\ell$ varies. As $M\ell$ increases, the modes 
on these  branches migrate away from the real axis.
The branch closest to the vertical axis is the most important dynamically 
because its imaginary part is small. 

This branch is connected to the origin
of the complex plane in the limit $M\ell\to0$ 
and the corresponding quasi-normal
modes have been computed analytically just above  for $H\ell\ll 1$ and
$M^2\ell^2\ll 1$.
%%%%%%%%%%%%%%%%%%%%%%%%%%%%%%%%%%%%%%%%%%%%%%%%%%%%%%%%%%%%%%%%%%%%%%
\begin{figure}
\begin{center}
\includegraphics[width=4.8in]{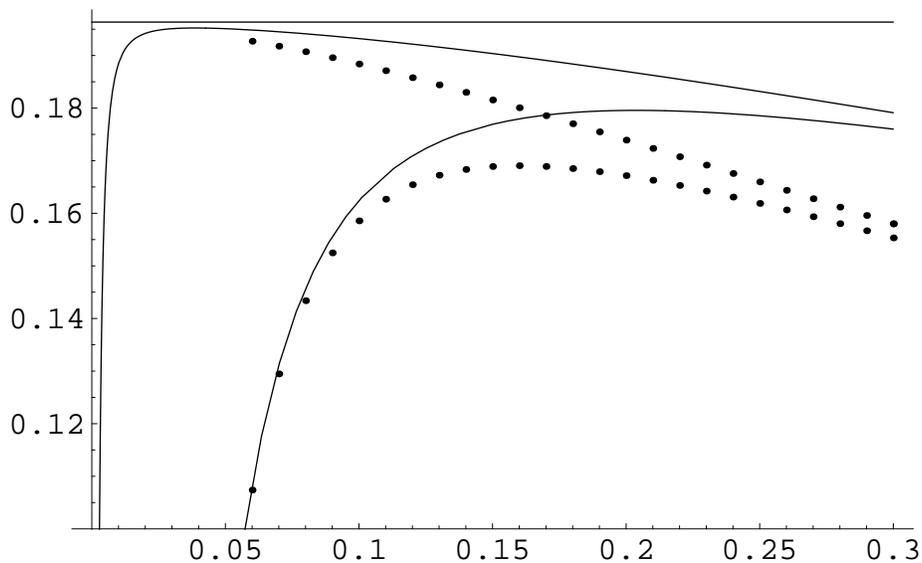}
\end{center}
\caption{$-\Im[m_{(4)}^2]/(M^4\ell^2)$ as a function of $M\ell$ ($z=50$ for 
the lower points, $z=1000$ for the upper points). The curves correspond 
to the analytical estimate (\ref{imaginary})}
\label{fig_im}
\end{figure}
%%%%%%%%%%%%%%%%%%%%%%%%%%%%%%%%%%%%%%%%%%%%%%%%%%%%%%%%%%%%%%%%%%%%%
As a check, we can compare the analytical results given 
above  with our numerical solutions. 
As noted before, we perform our search in the first quadrant of
the complex $\mu$-plane, which means ${\rm Im}[m_{(4)}^2]<0$.
 First, we have checked that the real part of $m_{(4)}^2$ is indeed 
very close  to the analytical value  
$M^2/2$. 
For the imaginary part, we have plotted in Fig. \ref{fig_im}
$-\Im[m_{(4)}^2]/(M^4\ell^2)$ as a function of $M\ell$
in two cases, $z=50$ and $z=1000$. 
In both cases, starting from high values of $M\ell$, 
 one approaches the analytical
value of $\pi/16$ as $M\ell$ decreases, as suggested by (\ref{mu2im}). 
But, whereas in the case $z=1000$, the 
dots continue to approach $\pi/16$, there is a sudden  change in the 
evolution for $z=50$.  This is simply due to the fact that, even if 
$H\ell\ll 1$, the small values of $M\ell$ become of the same order
of magnitude as $H\ell$ and (\ref{mu2im}) is no longer a good 
approximation: one needs the more general expression (\ref{imaginary}). 
 The two curves in the figure correspond 
to the analytical estimate given by (\ref{imaginary}) and one sees 
that the numerical values converge towards these curves for low masses.

Our numerical treatment allows us to continue this quasi-normal branch 
beyond the regime of validity of the analytical calculations,
and in particular for very high masses.  
Moreover, as mentioned before and as illustrated in Fig. \ref{qnm1}, 
we have found other solutions for $\mu$, 
away from the real axis, which correspond to additional
 branches. Each of this branch starts from the real axis after 
a critical mass threshold has been reached and then evolves away from 
the real axis as $M\ell$ increases.

%%%%%%%%%%%%%%%%%%%%%%%%%%%%%%%%%%%%%%%%%%%%%%%%%%%%%%%%%%%%%%%%%%%%%
\begin{figure}
\begin{center}
\includegraphics[width=4.8in]{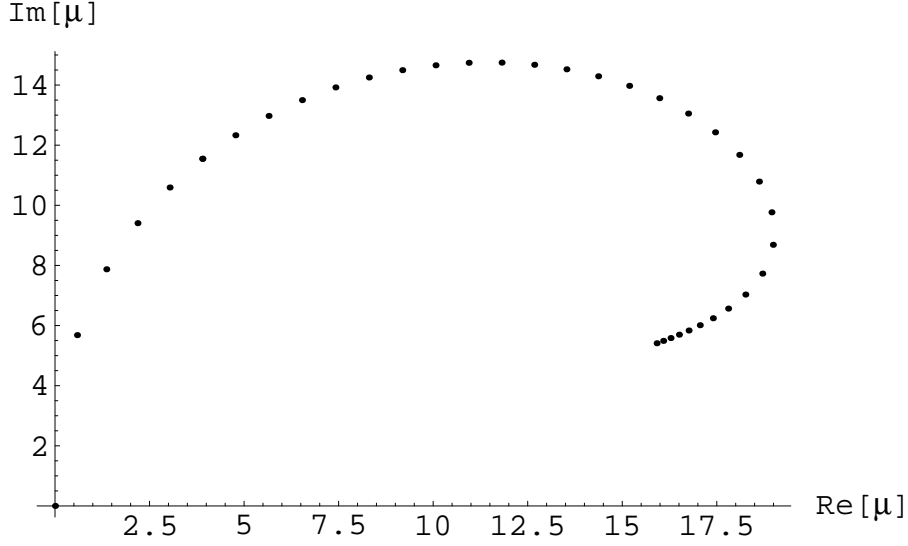}
\end{center}
\caption{Evolution in the complex $\mu$-plane of the quasi-normal
modes when the coupling varies,
in the clockwise direction, from $0$ to $7$
(with $M\ell=0$, $z=10$). The increment between two points is $\Delta\e=0.2$.
} 
\label{qnmC}
\end{figure}
%%%%%%%%%%%%%%%%%%%%%%%%%%%%%%%%%%%%%%%%%%%%%%%%%%%%%%%%%%%%%%%%%%%%%
\begin{figure}
\begin{center}
\includegraphics[width=4.8in]{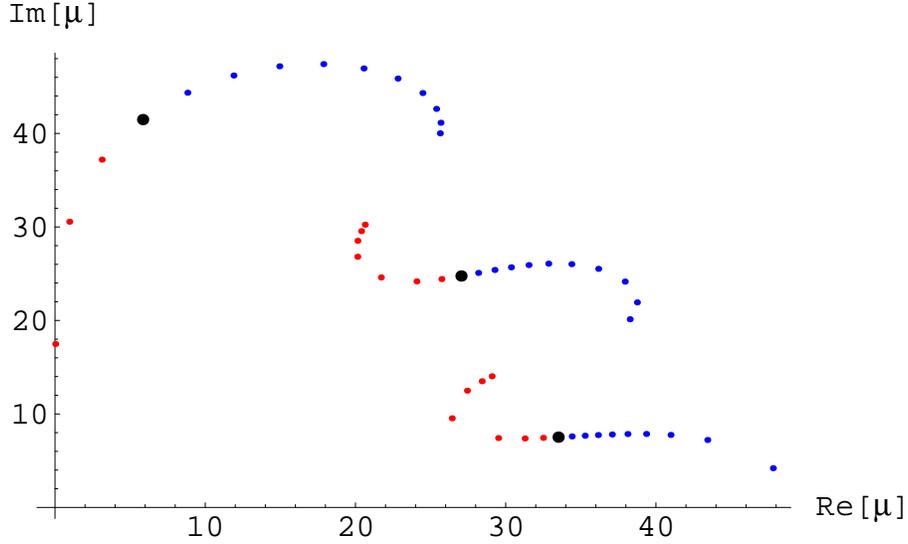}
\end{center}
\caption{Variation in the complex $\mu$-plane of the quasi-normal
modes when the coupling changes (for $M\ell=5$, $z=10$). The increment 
between adjacent points is $\Delta\e=1$. The three big dots (in black)
correspond to the modes for $\e=0$. The points on the left of 
each black dot (in red) correspond to 
negative couplings and those on the right (in blue) to positive couplings.}
\label{qnmC_M5}
\end{figure}
%%%%%%%%%%%%%%%%%%%%%%%%%%%%%%%%%%%%%%%%%%%%%%%%%%%%%%%%%%%%%%%%%%%%%
Let us now consider the situation when the coupling is allowed to vary.
 We have plotted in Fig.~\ref{qnmC}
the evolution of the quasi-normal mode with the increasing (positive) 
coupling in the case $M\ell=0$. What can be observed is that the mode,
 after going away from the real axis, tends to come 
back towards it.

  In Fig.~\ref{qnmC_M5}, we have also plotted
 the same evolution, i.e. with increasing coupling,  in the case 
$M\ell=5$ where several quasi-normal modes are already known to 
exist in the complex $\mu$-plane for vanishing coupling.
We have no analytic expression for this case, but 
one can observe  each of the modes follows a  curve analogous to the  
$M\ell=0$ case. We have also added the evolution corresponding to increasing 
negative coupling. For the first branch, one finds that the mode converges 
towards the origin of the complex plane, which could be expected since at some
critical value of the coupling one will enter into the localized mode region.
For the second branch, one observes that the modes at large negative 
couplings converge towards the modes of the previous branch at large 
positive couplings. The same behaviour occurs for the third branch.

\section{Conclusions}
In the present work, we have studied the modes of a massive bulk scalar field
that are localized or (quasi-localized) on a brane with a 
 de Sitter geometry,  characterized by a Hubble parameter $H$. We have allowed
for a coupling between the brane tension and the bulk scalar field, 
quantified by a dimensionless parameter $\e$. 

Although the bound state of a massless bulk scalar field without coupling 
to the de Sitter brane is a zero mode, i.e. its four-dimensional 
effective mass $m_{(4)}$ vanishes, this is no longer the case 
for a non-zero bulk mass $M$ or a non zero coupling. One thus finds 
in general a {\it massive bound state}, whose four-dimensional mass
depends on $M$, $\e$ and $H$. We have computed explicitly this 
dependence in limit of a small bulk mass and small coupling. 

The mass-squared $m_{(4)}^2$ of the bound state is always comprised between 
zero and $(3/2)H^2$, value above which one finds a continuum of modes. 
We have obtained explicitly, for any value of $M$ and $H$, the lower 
and upper critical values for the coupling corresponding to this range 
of possible zero modes. 

Another approach consists in trying to find a purely four-dimensional 
description of the scalar field including the effect of the bulk. We have 
generalized previous such procedures to take into account the coupling, and 
defined an effective potential for the value on the brane of the scalar field. 
For small $M$ and $\e$, the second derivative of the effective 
potential yields an effective mass-squared which is in excellent agreement 
with the rigorous mode approach for small $H$ and in qualitative agreement
for large $H$. We have also demonstrated that a class of known exact solutions
with a bulk scalar field indeed conforms to the effective potential
approach. 

Finally, when the three parameters  $M$, $\e$ and $H$ are not 
in the region allowing for bound states, one finds modes whose effective 
mass has an imaginary part. This means that the corresponding states 
cannot stay localized on the brane but will escape into the bulk. We have 
explored numerically the dependence of the real and imaginary part of the 
four-dimensional mass on the values of the bulk mass and of the coupling and
we have given an analytical estimate of the quasi-localized  modes 
for small values of the Hubble parameter, i.e. $H\ell\ll 1$. 

\acknowledgements
We would like to thank K. Koyama for useful communications,
in particular, for bringing our attention to exact solutions
that conform to the effective potential approach.
This work is supported in part by Monbukagaku-sho Grant-in-Aid
for Scientific Research (S), No.~14102004. This work was initiated
when MS was visiting the gravitation and cosmology group (GReCO)
at IAP, Paris. MS is grateful to the Universit\'e Paris 7-APC 
for financial support and to the GReCO members for warm
hospitality.

\end{document}